\documentclass[11pt]{article}

\usepackage[final]{acl}

\usepackage{multirow}
\usepackage{colortbl}
\usepackage[most]{tcolorbox}
\usepackage{lipsum}
\usepackage{booktabs}
\tcbuselibrary{breakable}

\usepackage{times}
\usepackage{latexsym}
\usepackage{listings}
\usepackage{xcolor}

\lstdefinelanguage{json}{
    basicstyle=\small\ttfamily,
    numbers=left,
    numberstyle=\ tiny,
    stepnumber=1,
    numbersep=8pt,
    showstringspaces=false,
    breaklines=true,
    frame=single,
    backgroundcolor=\color{lightgray!20},
    string=[s]{"}{"},
    comment=[l]{:\ "},
    morecomment=[l]{:"},
    literate=
        *{0}{{{\color{blue}0}}}{1}
         {1}{{{\color{blue}1}}}{1}
         {2}{{{\color{blue}2}}}{1}
         {3}{{{\color{blue}3}}}{1}
         {4}{{{\color{blue}4}}}{1}
         {5}{{{\color{blue}5}}}{1}
         {6}{{{\color{blue}6}}}{1}
         {7}{{{\color{blue}7}}}{1}
         {8}{{{\color{blue}8}}}{1}
         {9}{{{\color{blue}9}}}{1}
}

\usepackage[T1]{fontenc}

\usepackage[utf8]{inputenc}

\usepackage{microtype}

\usepackage{inconsolata}
\usepackage{amssymb}

\usepackage{graphicx}

\title{SmellBench: Towards Fine-Grained Evaluation of Code Agents on Refactoring Tasks}

\author{
Fake Lin$^{1,2}$,
Binbin Hu$^{2}$,
Xi Zhu$^{3}$,
Ziwei Zhao$^{1}$,
Zhi Zheng$^{1}$\\
\textbf{Ziqi Liu}$^{2}$,
\textbf{Zhiqiang Zhang}$^{2}$,
\textbf{Jun Zhou}$^{2}$,
\textbf{Tong Xu}$^{1}$\\[0.5em]
$^{1}$University of Science and Technology of China \quad
$^{2}$Ant Group \quad $^{3}$Rutgers Univerisity\\[0.3em]
fklin@mail.ustc.edu.cn, bin.hbb@antfin.com, zhengzhi97@ustc.edu.cn\\[0.3em]
\raisebox{-0.2em}{\includegraphics[height=1.0em]{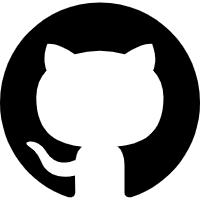}} \small{Code: \texttt{\href{https://github.com/MINE-USTC/SmellBench}{https://github.com/MINE-USTC/SmellBench}}} \\
\raisebox{-0.2em}{\includegraphics[height=1.0em] {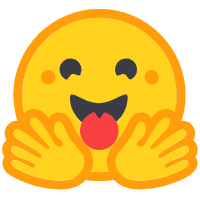}} \small{Data: \texttt{\href{https://huggingface.co/datasets/critical88/SmellBench}{https://huggingface.co/datasets/critical88/SmellBench}}}
}

\begin{document}
\maketitle
\begin{abstract}


Code Agents have achieved remarkable advances in recent years, exhibiting strong capabilities across a wide range of software engineering tasks. However, their misuse often produces bloated and disorganized code that impairing readability, extensibility, and robustness. 
Despite this risk, existing benchmarks largely evaluate functional correctness rather than long-term maintainability of code agents.
In this paper, we propose SmellBench, an extensible code refactoring benchmark that proactively injects code smells into clean code snippets from real-world repositories. This design enables the generation of controlled, high-quality, and diverse refactoring cases with human-written ground truth. Specifically, it contains 294 cases spanning 7 popular smell types, 3 difficulty levels, 2 instruction settings across 7 real-world repositories. We further design 3 evaluation aspects covering functional correctness, localization ability, and refactoring quality assessment.
Experiments with 2 popular agents and 6 large langauge models (LLMs) show that the best combination—Qwen Code + Claude Sonnet 4.5—achieved only a 50.34 score of smell elimination. Further analysis reveals that this gap arises from a focus on local code smells and a lack of cross-file understanding, which hinders comprehensive smell elimination.
\end{abstract}

\section{Introduction}
Recent advances in large language models (LLMs) have sparked the development of powerful code agents capable of solving complex software engineering tasks, including code generation, bug fixing, and repository-level reasoning. \cite{wang2024openhands,yang2025qwen3,yang2024swe,aider2023aider}. While bringing convenience to developers, many programming tasks have been delegated entirely to code agents without any human review. Such misuse raises broad concerns about the generation of bloated and disorganized code, which severely impairing readability, extensibility, and robustness. 
However, existing benchmarks primarily focus on functional correctness, leaving the long-term maintainability of code agents largely underexplored.
\cite{jimenez2023swe,merrill2026terminalbenchbenchmarkingagentshard,jain2024livecodebench,zan2025multiswebench}. 
Fortunately, code refactoring has garnered increasing attention, where the task aims at improving existing code quality, enhancing maintainability, and supporting the long-term evolution of software systems \cite{liu2025exploring,gautam2025refactorbench,garg2025perfbench,dinu2026smellbenchevaluatingllmagents}. 

Existing refactoring benchmarks can be broadly divided into two lines. One line of research focuses on curating refactoring-oriented benchmarks \cite{chen2021evaluating,gautam2025refactorbench,jain2024livecodebench,tapader2025code}, which offer high-quality refactoring cases with clear instructions. However, the substantial human effort leads to unaffordable cost and limited scalability, which in turn restricts the coverage of diverse refactoring scenarios in real-world codebases. Another line of benchmark automatically mines refactoring cases from the commit histories of GitHub repositories\cite{xu2026swerefactorrepositorylevelbenchmarkrealworld,pomian2024next,kovacic2025refactoring,DBLP:journals/corr/abs-2603-04177}. Although this approach offers better scalability, real-world refactoring cases are often distributed across multiple commits and tightly intertwined with feature implementation and bug fixing. Consequently, refactoring signals in real repositories are sparse, fragmented, and unevenly distributed, making it difficult for commit-history-based benchmarks to systematically capture realistic refactoring scenarios.

To address this limitation, we propose \textbf{SmellBench}, an extensible code refactoring benchmark that \emph{proactively injects
code smells into clean code snippets from real-world repositories}. This design supports the generation of controlled, high-quality, and diverse refactoring cases with human-written ground truth. 
SmellBench contains 294 cases spanning 7 smell types, 3 difficulty levels and two instruction settings across 7 real-world repositories to ensure comprehensive evaluation of refactor ability of current code agents. 
Overall, our SmellBench offers the following advantages: (1) \textbf{Purity}. It isolates refactoring from other code changes. (2) \textbf{Diversity}. It covers 7 complex smell types.
(3) \textbf{Authenticity}. It provides human-written ground truth. (4) \textbf{Extensibility}. It can be easily extended to other smell types or even task types by introducing new specifications. (5) \textbf{Decontamination}. It avoids data leakage as it only involves newly generated cases.

We further design 3 evaluation aspects specifically for refactoring tasks, including \textbf{test passing rate} to ensure functional correctness, \textbf{localization accuracy} to assess the identification of refactoring targets, and \textbf{LLM-based judgment} to evaluate the refactoring quality. Moreover, we conduct experiments on 2 open-source code agents and 6 representative LLMs. The results show that the best-performing configuration, Qwen Code with Claude Sonnet 4.5, achieves only 50.34 score of smell elimination, which represents that existing approaches mainly focus on local code smells and struggle to identify and address smells that propagate across multiple files. Meanwhile, most models pass over 80\% test cases, suggesting that relying solely on functional correctness is insufficient to effectively distinguish refactoring capabilities among code agents. This further highlights the necessity of introducing LLM-as-Judge for refactoring quality assessment.

In summary, we highlight the main strengths of this work as follows:
\begin{itemize}
    \item We propose \textbf{SmellBench}, an extensible code refactoring benchmark that proactively injects code smells into clean code snippets, achieving the trade-off between data scalability and quality in existing benchmarks.
    
    \item SmellBench comprises 294 high-quality refactoring cases spanning 7 complex smell types, 3 difficulty levels, and 2 instruction settings across 7 real-world repositories, with human-written ground truth for evaluation.
    
    \item We design a 3-dimensional evaluation framework tailored for refactoring tasks, including test passing rate, localization accuracy, and LLM-based quality assessment, providing more nuanced insights beyond functional correctness.
    
    \item Extensive experiments with 2 open-source agents and 6 representative LLMs reveal that current approaches achieve only 50.34 score in smell elimination, struggling particularly with cross-file smell identification, which revealing significant room for improvement.
\end{itemize}

\section{Related Work}

\subsection{Code Benchmark}
Benchmarking the coding capabilities of large language models has evolved with the increasing complexity of code generation tasks \cite{chen2021evaluating,austin2021program,jimenez2023swe,jain2024livecodebench,ni2025gittaskbench,zhuo2024bigcodebench}. Early benchmarks such as HumanEval \cite{chen2021evaluating} and MBPP \cite{austin2021program} focus on isolated programming problems and measure the functional correctness of generated code at the function or algorithm level.
Beyond standalone functions, ClassEval \cite{du2023classeval} targets class-level code generation, and repository-oriented benchmarks including RepoBench \cite{liu2023repobench} and CrossCodeEval \cite{ding2023crosscodeeval} extend the scope to cross-file and project-level code modeling, emphasizing contextual reasoning in realistic codebases.
Subsequently, research attention shifted toward real-world software engineering problems. SWE-bench \cite{jimenez2023swe} centers on bug fixing derived from real GitHub issues. LiveCodeBench \cite{jain2024livecodebench} draws tasks from contests across three competitive programming platforms, placing greater demands on existing large language models.
More recently, task-oriented benchmarks have been proposed to assess code agents under realistic user demands. GitTaskBench \cite{ni2025gittaskbench} anchors benchmarking in repository-based tasks, and CodeIF \cite{yan2025codeif} characterizes instruction-following behavior across diverse code generation scenarios, shifting the paradigm toward user-driven, multi-step problem solving.

However, these benchmarks focus mainly on code generation and bug fixing, but pay limited attention to refactoring tasks that are fundamental and critical to software maintenance and long-term evolution.

\begin{figure*}[htbp]
  \centering
  
  \includegraphics[width=\linewidth]
  {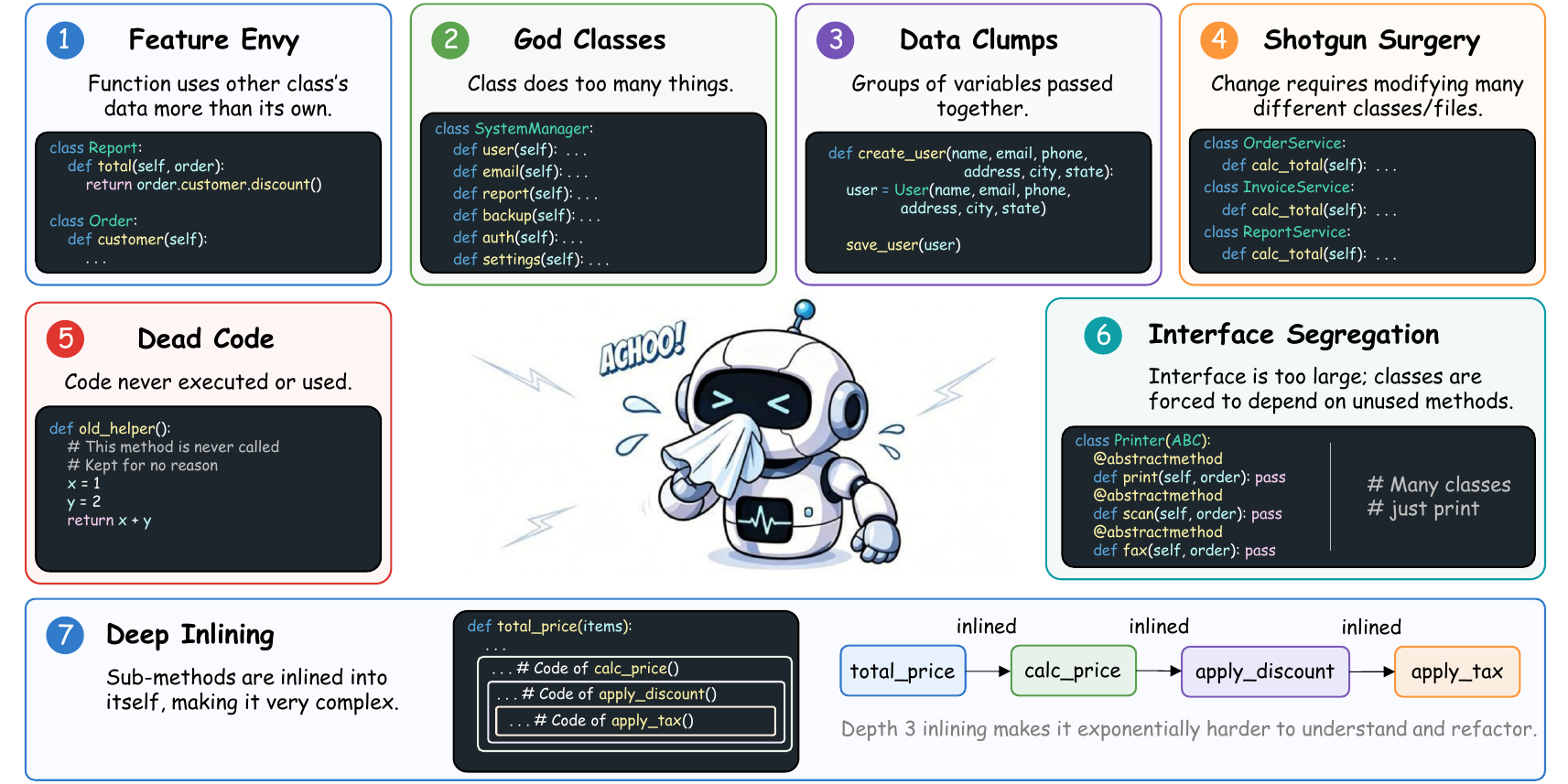}
  \caption{ Definitions of the 7 code smell types with illustrative toy examples. }
  \label{fig:smell}
  
\end{figure*}

\begin{figure}[htbp]
  \centering
  \includegraphics[width=\linewidth]
   {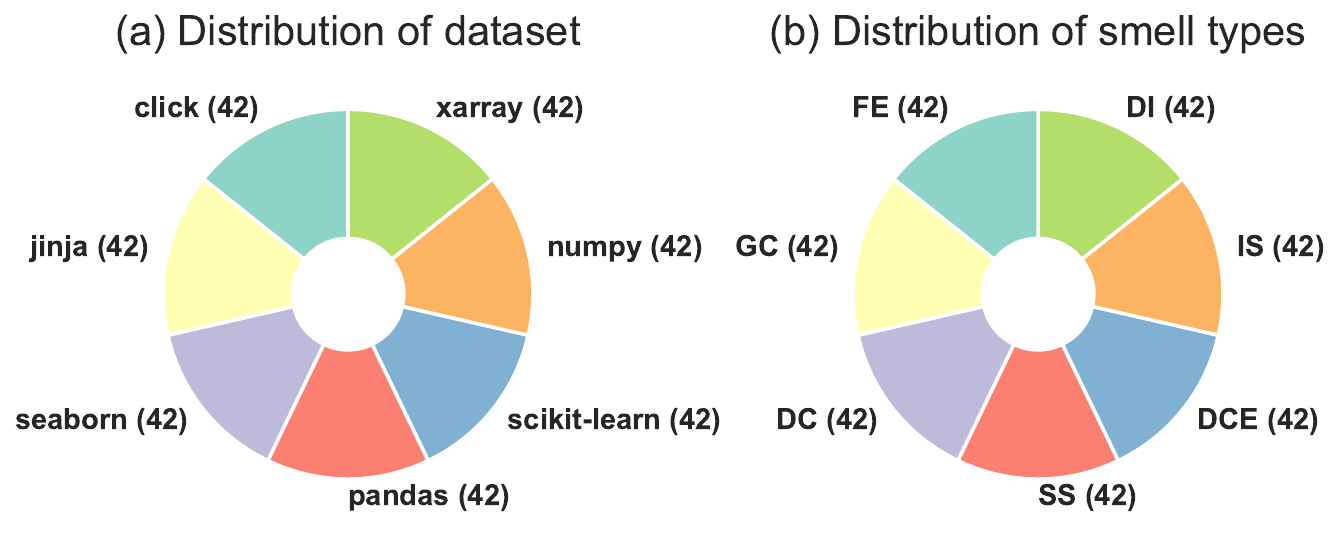}
  \caption{Distribution of SmellBench across 7 python repositories and 7 smell types.}
  \label{fig:dataset_dist}
  
\end{figure}
\subsection{Refactor Benchmark}
Alongside the rapid progress of code generation benchmarks, research on evaluating code refactoring has gained increasing attention in recent years \cite{gautam2025refactorbench,xu2026swerefactorrepositorylevelbenchmarkrealworld,liu2025exploring,kuang2025refactoring}. 
Early work constructs refactoring datasets from a small number of projects, mostly in Java, and evaluates LLMs on identifying refactoring opportunities and applying corresponding code changes \cite{liu2025exploring,tapader2025code}. These studies show that prompt design can significantly affect refactoring performance, but their primary goal is to analyze LLM behavior instead of modeling realistic refactoring workloads.  
Later efforts adopt more complex settings for refactoring, such as stateful refactoring across multiple steps \cite{gautam2025refactorbench}, extracting shared logic into reusable libraries \cite{kovacic2025refactoring}, or combining LLM suggestions with IDE or static analysis tools \cite{pomian2024next}. These directions emphasize broader code context and design-level reasoning.
More recent benchmarks attempt to leverage real-world repositories by mining refactoring-related commits or covering a wider range of refactoring types \cite{kuang2025refactoring,xu2026swerefactorrepositorylevelbenchmarkrealworld}. However, these approaches often rely heavily on external refactoring detection tools or curated examples, limiting scalability and language coverage.
Concurrent work, CodeTaste \cite{DBLP:journals/corr/abs-2603-04177}, extends commit-based benchmarks to multilingual and multi-file refactoring tasks, resulting in more complex and realistic evaluation settings that better reflect real-world software development practices.

In summary, existing refactoring benchmarks either rely on manually constructed datasets that are costly and limited in scale, or mine refactoring instances from repository commits that inevitably mix refactoring with other code changes, leading to limited coverage, scalability, or data purity.

\begin{figure*}[htbp]
  \centering
  \includegraphics[width=\linewidth]
  {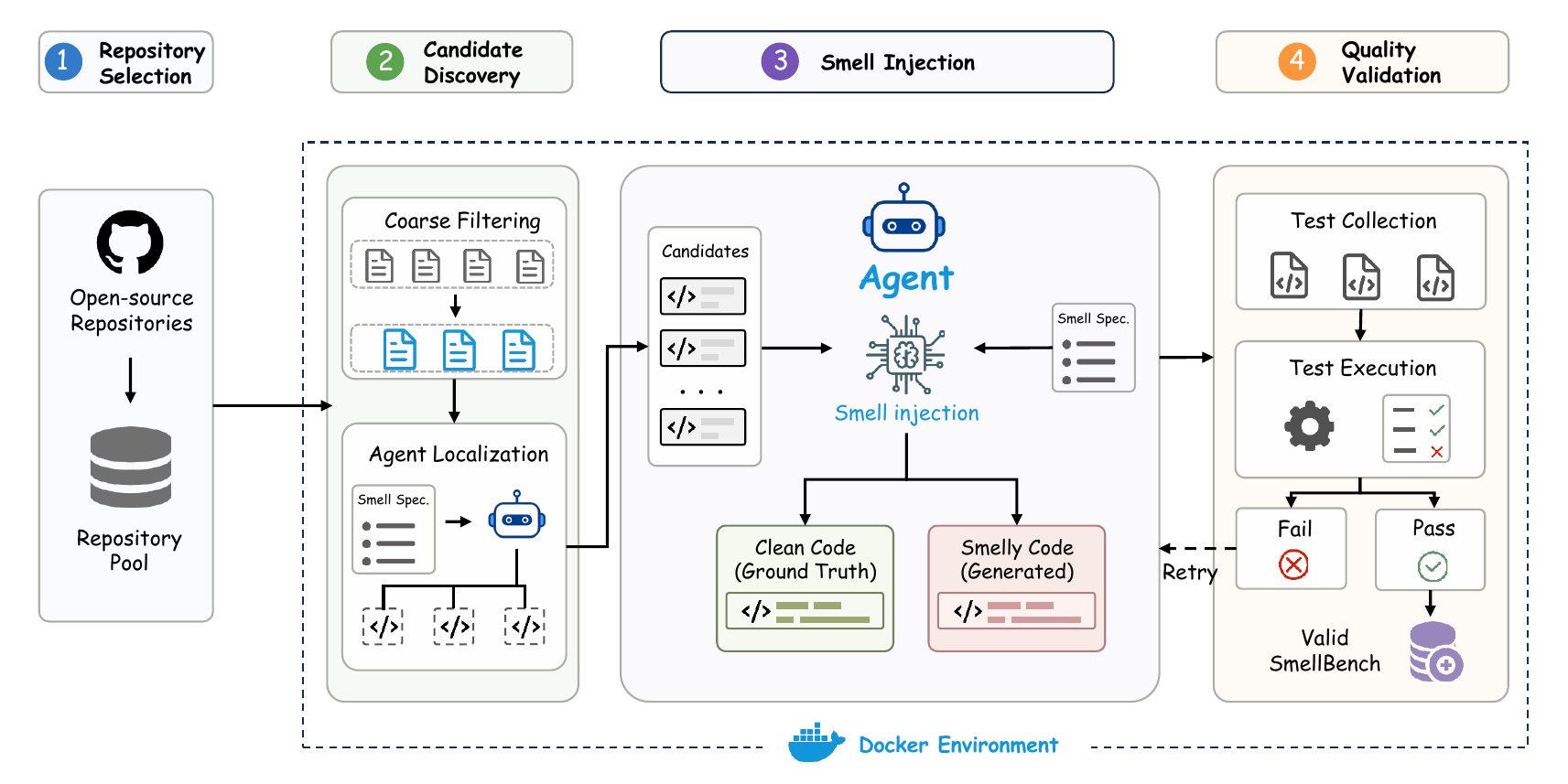}
  \caption{SmellBench construction pipeline: Repository Selection to collect repositories, Candidate Discovery to identify suitable injection locations, Smell Injection to introduce controlled code smells, and Quality Validation to validate correctness and benchmark reliability. The latter three stages are executed in a Dockerized environment. }
  \label{fig:pipeline}
  
\end{figure*}
\section{SmellBench}
This section is organized as follows: First, we provide a brief description of the 7 smell types. Next, we present the construction pipeline of the benchmark, which consists of four steps. Finally, we describe the evaluation process and the definition of 3 metrics.

\subsection{Smell Types}


Code smells are a long-established abstraction in software engineering for characterizing structural design deficiencies that motivate refactoring. Importantly, they naturally encode refactoring objectives that go beyond local edits, often involving interactions across functions, classes, and files. By grounding refactoring tasks in a curated set of representative code smell types, we can construct refactoring scenarios that are both structurally meaningful and amenable to controlled evaluation.

Specifically, we instantiate refactoring tasks using 7 representative smell types \cite{Fowler1999}. Their formal definitions refer to Appendix \ref{sec:smell_types} and illustrative toy examples are provided in Figure \ref{fig:smell}.

\subsection{Benchmark Construction}
\label{sec:benchmark_construction}
The construction of SmellBench follows four steps: collecting real-world software repositories, identifying suitable injection locations, introducing code smells into clean code, and validating functional correctness using test cases. The overall workflow is illustrated in Figure \ref{fig:pipeline}.

\subsubsection{Repository Selection} We select 7 Python repositories from top-ranked PyPI projects as our codebases. Popular, large-scale repositories are preferred due to their mature contributor guidelines, more comprehensive test coverage, and complex code dependencies.

\subsubsection{Candidate Discovery}
Following repository collection, we identify appropriate locations for injecting specific code smells. This stage is designed to ensure both coverage and validity of injection locations.
First, real-world software repositories are typically large and heterogeneous. Allowing a code agent to search for injection locations without constraints can lead to location bias, where the agent disproportionately selects candidates from early-accessed files while neglecting other suitable regions of the codebase.
Second, code smells are not universally applicable. Each smell type imposes distinct structural or semantic requirements, meaning that only a subset of functions or classes constitute valid injection locations. Consequently, candidate selection must be guided by smell-specific constraints rather than unrestricted search. Along this line, Candidate Discovery is decomposed into two submodules:

\paragraph{Coarse Filtering.} This module reduces the search space by pre-selecting files that are more likely to contain meaningful injection targets. In our current setup, we retain source files exceeding 500 lines of code, which prioritizes files with sufficient structural richness for multi-file refactoring scenarios.

\paragraph{Agent Localization.} Given the filtered files, a code agent is used to identify candidates that satisfy the requirements of the smell specification. The code agent excels at program comprehension and contextual reasoning to analyze code semantics, dependencies, and structural patterns, ensuring more reliable identification of smell injection candidates than rule-based matching approaches. Notably, This step focuses exclusively on locating and selecting candidates and does not perform any code modifications. The resulting candidate set is passed to the subsequent smell injection stage.
\subsubsection{Smell Injection}
After obtaining clean code candidates, this stage automatically constructs code samples containing target code smells. Specifically, the code agent is fed with candidates and the smell specification, which directly modifies the original code to introduce the target smell while preserving compilability.

In contrast to prior methods, this stage relies solely on a sufficiently powerful code agent (e.g., OpenHands + Claude Sonnet 4.5), and therefore frees from programming language, external tools or any commit histories. As a result, SmellBench is theoretically capable of producing unlimited number of cases via injecting various smells across different candidates, which overcomes the challenges of scalability and diversity arising from commit-based approaches.

To facilitate a systematic analysis of agent behavior, we further introduce 3 levels of case difficulty and 2 types of instruction settings. These configurations allow us to examine how code agents respond to varying levels of task complexity and instructional constraints. Detailed information of these settings are provided in Appendix \ref{sec:difficulty} and \ref{sec:instruction}.

Finally, we retain both the original and modified versions of the code, forming paired clean–smelly samples that serve as the basis for subsequent quality validation and evaluation.

\subsubsection{Quality Validation}
Recall that refactoring tasks aim to improve code structure while preserving program semantics and utility. Therefore, we are supposed to ensure functional correctness of smelly code, i.e., the created smelly code must pass the unit tests. 

To this end, we collect test suites related to the smelly code and execute the associated tests. Specifically, coverage analysis is employed to determine which test cases cover the methods contained in the smelly code. Each smelly instance include several methods and each method is associated with at most 10 relevant test cases. Test-failing cases, along with their error messages, are fed back to the code agent in Smell Injection stage for rectification to enhance both pass rates and generation efficiency. Finally, cases are filtered out if no relevant tests available or if the tests fail again. 

This test-based validation process allows SmellBench to retain high-quality and non-trivial refactoring cases, supporting a comprehensive assessment of code agent refactoring capabilities.

After these four steps, we finally generate 294 valid samples with 7 smell types, 3 difficulty levels and 2 instruction settings across 7 repositories. The statistics of the SmellBench are illustrated in Figure \ref{fig:dataset_dist}, where each category is evenly distributed, demonstrating the controllability of the construction pipeline.

\begin{table*}[h!]
\centering
\caption{Performance comparison of different agents and LLMs on SmellBench. Oracle means the ground truth, TPR represents the Test Pass Rate, LA denotes the Localization Accuracy, CQ is code quality, SS stands for structural soundness, CC refers to cross-file coordination, SE indicates smell elimination. The bold numbers indicate the best results.}
\label{tab:overall}
\begin{tabular}{cccccccc}
\toprule
\multirow{2}{*}{Agent} 
& \multirow{2}{*}{LLM}
& \multirow{2}{*}{TPR}
& \multirow{2}{*}{LA}
& \multicolumn{4}{c}{LLM-as-Judge}\\

\cmidrule(lr){5-8}

& & & 
& CQ
& SS
& CC
& SE
\\

\midrule
\midrule
Oracle &	- &	1.0 &	1.0 & 	0.9592	& 0.9626 &	0.9884 & 0.9660 \\
\hline
\multirow{6}{*}{OpenHands} & Qwen-Coder-30B-A3B & 0.8021 & 0.7067 &	0.4649 &	0.4110 & 0.3014 &  0.3004  \\
& Qwen-Coder-480B-A35B & 0.8258 & 0.7456 &	0.5647 & 0.5084 &	0.3962 & 0.4080 \\
& DeepSeek-V3.2 & 0.8790 &	0.7224 &	0.5801 & 0.5203 &	0.3804 &	0.3733  \\
& Gemini-2.5-Flash & 0.5272 &	0.6667 & 	0.4041 & 0.3592 &	0.2323 & 0.2327 \\
& GPT-5-Mini & 0.9286 &	0.7721 &	0.6146 &	0.5418 & 	0.4316 &	0.4235 \\
& Claude-Sonnet-4.5 & 0.9116	& 0.8401	& 0.6779 &	0.6139 &	0.4881 &	0.4793 \\
\midrule
\multirow{6}{*}{Qwen Code} & Qwen-Coder-30B-A3B & 0.7789 &	0.6701 &	0.4203 &	0.3574 &	0.2344 &	0.2251  \\
& Qwen-Coder-480B-A35B & 0.8197 &	0.6531 &	0.4573 &	0.3990 &	0.2812 &	0.2867 \\
& DeepSeek-V3.2 & 0.8571 &	0.4966 &	0.3677	& 0.3248	& 0.2405 &	0.2503 \\
& Gemini-2.5-Flash & 0.6633 &	0.6395 &	0.4163 &	0.3735 &	0.2231	& 0.2279 \\
& GPT-5-Mini & 0.9218	& 0.7483	& 0.5626	& 0.4871 &	0.3238 &	0.3255  \\
& Claude-Sonnet-4.5 & \textbf{0.9388} &	\textbf{0.8878} & \textbf{0.6929} & \textbf{0.6405} &	\textbf{0.5143} &	\textbf{0.5034} \\
\bottomrule
\end{tabular}
\end{table*}

\subsection{Evaluation Framework}
To comprehensively assess code agents in refactoring tasks, we examine not only whether their changes pass existing tests but also the overall quality of the refactoring in terms of structural soundness and smell elimination. Specifically, our evaluation consists of three aspects, including test passing rate to ensure functional correctness, localization accuracy to assess the identification of refactoring targets, and LLM-based judgment to evaluate the refactoring quality.

\paragraph{Test Pass Rate.} Test pass rate (TPR) measures whether the code refactors preserve the original program behavior. A refactoring is considered successful if the refactored program passes the entire test suite. Otherwise, it is counted as a failure.

\paragraph{Localization Accuracy.} Localization accuracy (LA) evaluates whether a code agent correctly identifies the code locations that require refactoring. Depending on the smell type, localization is evaluated at either the class level or the method level. A prediction is considered correct if the identified class or method matches the ground-truth refactoring target.

\paragraph{LLM-as-Judge.} LLM-as-Judge employs LLMs to assess refactoring quality from four complementary aspects: \emph{code quality} (CQ), which measures readability and maintainability improvements; \emph{structural soundness} (SS), which evaluates the correctness and rationality of the refactored structure; \emph{cross-file coordination} (CC), which assesses consistency and dependency handling across multiple files; and \emph{smell elimination} (SE), which measures the effectiveness of removing the targeted code smells.
The detailed rubric information can be found in Appendix \ref{sec:rubrics}.

In LLM-as-Judge part, the LLM conducts a comparative analysis between the code smells and the refactoring results from agent, and derives an evaluation judgment. However, this process demands strong contextual understanding and reasoning capabilities from the LLM to disentangle the structural issues and logical relationships implicitly embedded in complex code smells, which can easily lead to inconsistent evaluation outcomes across different LLMs. To enable stable and consistent conclusions among different models, we introduce a preliminary step, Smell Analysis.

\paragraph{Smell Analysis.} Smell analysis explicitly analyzes and summarizes the code smells to clarify the corresponding structural issues and refactoring objectives, and provides clear and unified contextual grounding for subsequent LLM-based evaluation. Please refer to Figure \ref{fig:smell_analysis_example} for examples.

Notably, SmellBench has been integrated with the Harbor framework \cite{Harbor_Framework}, which performs evaluation within Docker environments and supports a wide range of popular code agents.

\section{Experiments}

\subsection{Experimental Settings}
\paragraph{Model.} To examine whether existing code agents can effectively address refactoring issues, we select some of the most popular and powerful agents together with representative LLMs. Specifically, we choose two widely used open‑source agents, OpenHands \cite{wang2024openhands} and Qwen Code \cite{yang2025qwen3}, paired with six large language models (LLMs), including three cutting‑edge open‑source models — Qwen3‑Coder‑30B‑A3B‑Instruct \cite{yang2025qwen3}, Qwen3‑Coder‑480B‑A35B‑Instruct \cite{yang2025qwen3}, and DeepSeek‑V3.2 \cite{liu2025deepseek} — and three closed-source models — GPT-5‑Mini \cite{achiam2023gpt}, Gemini-2.5‑Flash \cite{comanici2025gemini}, and Claude Sonnet‑4.5 \cite{anthropic_claude_model_report_2025}. 
\paragraph{Settings.} All experiments are conducted using the Harbor framework in a Docker-based isolated environment. Agents are granted full permission to freely modify the codebase without human approval. We evaluate two instruction settings, including a \textit{guided} setting with only involved files and smell types and a \textit{targeted} setting with additional involved methods and classes. Each task is executed with a maximum time budget of 20 minutes, and due to computational cost, each configuration is run only once.

\begin{figure*}[htbp]
  \centering
  
  \includegraphics[width=\linewidth]
  {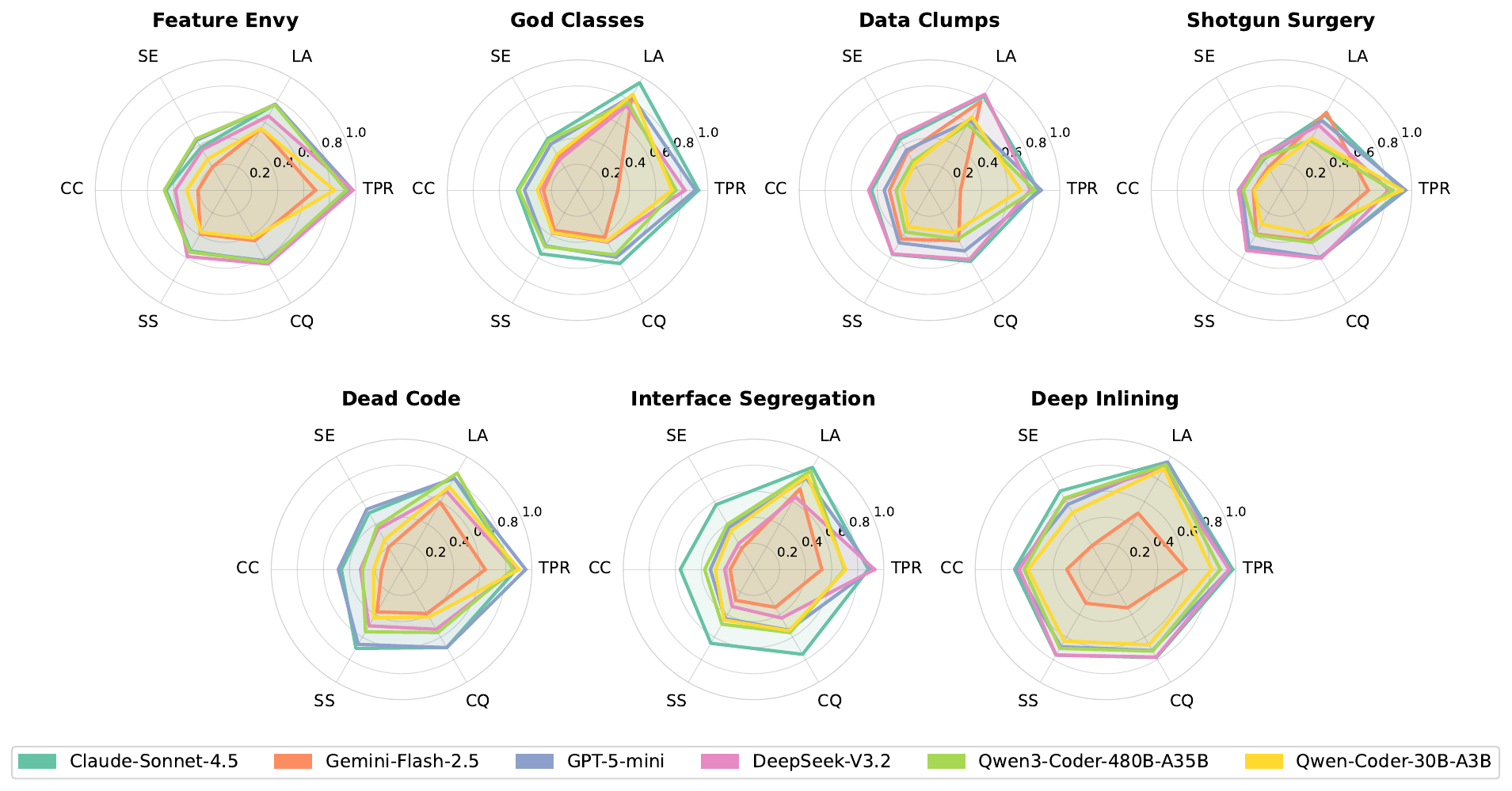}
  \caption{Performance comparison of 7 smell types under different LLMs with the OpenHands Agent. 6 metrics are adopted for comprehensive evaluation. }
  \label{fig:smell_type_ana}

\end{figure*}



\subsection{Overall Performance}
We report the overall performance in Table \ref{tab:overall}, which presents the code refactoring capabilities evaluated across two open-source agents and six large language models (LLMs). The Oracle row represents the original clean code before smell injection, achieving 0.96–0.99 scores rather than perfect 1.0 due to the LLM judge’s inherent conservatism in applying rubrics. This provides a calibration reference for interpreting absolute scores. From the results, we obtain the following key observations. (1) \textbf{Refactoring task is still challenging for SOTA baselines}. Even the best-performing baseline, QwenCode + Claude Sonnet 4.5, only achieved a Smell Elimination (SE) score of 0.5034 (see Appendix \ref{sec:rubrics} for details) , highlighting the significant gap between current code generation capabilities and the deeper structural reasoning required for high-quality refactoring. (2) \textbf{Test Passing Rate is not suitable for refactoring task.} Except for Gemini 2.5 Flash, almost all models achieve TPRs above 80\%, indicating that current models place strong emphasis on preserving functional correctness. In addition, the scores for Code Quality and Structural Soundness (SS) are also relatively high, further reinforcing this observation.
(3) \textbf{Open‑source models also exhibit strong competitiveness in code refactoring tasks.}
    Qwen-Coder-480B-A35B achieves a smell elimination score of 0.4080 and 0.2867 with OpenHands and QwenCode, respectively,  comparable to GPT-5-Mini (0.4235 and 0.3255, respectively). This demonstrates the rapid advancement of open‑source LLMs and their potential to rival closed-source models in complex code
    understanding and refactoring tasks.
(4) \textbf{OpenHands demonstrates better performance than Qwen Code when coupled with weaker LLMs}. However, as the LLM capacity increases, the performance gap between the two agents narrows. This suggests that OpenHands possesses stronger exploration capabilities, enabling more effective utilization of limited model capacity. When the LLM becomes more powerful, the upper bound of performance is constrained by the model’s inherent ability, leading both agents to converge toward similar results.

\subsection{Analysis of Smell Types}
In this subsection, we further analyze the performance of the seven code smells under different model configurations. The experiments employ OpenHands as the agent framework and select six representative large language models as backends. We evaluate the results using six metrics to reveal the differences in how various LLMs handle structural code issues. Based on the results shown in Figure \ref{fig:smell_type_ana}, we make the following observations:
(1) \textbf{Different code smells exhibit significant performance variations across different LLMs.}
Model rankings vary substantially across smell types. DeepSeek-V3.2 excels on Data Clumps and Deep Inlining, while performing less effectively on Interface Segregation, suggesting stronger structural reasoning than architectural design capabilities. GPT-5-Mini consistently achieves high test pass rates but is less competitive on design-oriented refactorings. These variations indicate that different smells emphasize different capabilities rather than a single notion of refactoring ability.
\begin{figure}[t]
  \centering
\includegraphics[width=\linewidth]
  {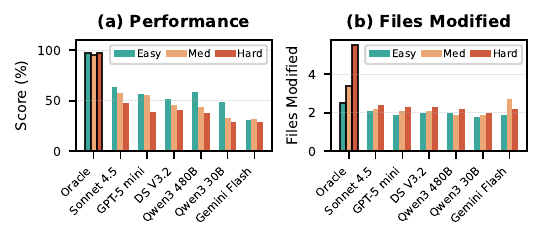}
  \caption{Performance comparison across difficulty levels. (a) scores (average of CQ, SS, CF, and SE) for Oracle and six models on Easy, Medium, and Hard tasks. 
  (b) Number of files modified by each model. }
  \label{fig:difficulty}
\end{figure}
(2) \textbf{All models perform poorly on Shotgun Surgery}.
Fixing Shotgun Surgery requires coordinated modifications across multiple files. Statistics show that this type of smell involves an average of 5.7 files, which is significantly higher than the overall average of 2.5 files. However, Claude Sonnet 4.5 modifies only 2.7 files on average when addressing this smell. This suggests that current approaches still have substantial limitations in handling multi-file coordinated modifications.
(3) \textbf{All models achieve strong performance on Deep Inlining.}
Repairing this type of smell mainly relies on the model’s ability to decompose code and perform deep reasoning. The experimental results indicate that all models perform well in these aspects, enabling them to effectively address this code smell.

\subsection{Analysis of Difficulty Levels}

This section mainly analyzes the impact of different difficulty levels on the number of task instances and model performance. By comparing the three difficulty levels—Easy, Medium, and Hard—we investigate the relationship between task difficulty and model capability. All experiments are conducted using OpenHands as the agent framework.

Based on the results shown in Figure \ref{fig:difficulty}, we draw the following conclusions:
(1) As shown in Figure \ref{fig:difficulty}(a), the performance of all models gradually declines as the difficulty level increases, demonstrating the rationality of the benchmark’s difficulty design. 
(2) Open-source models perform well on easy tasks but show relatively poor performance on difficult tasks, indicating that their deep reasoning capabilities still lag behind strong models such as Claude Sonnet 4.5.
(3) Figure \ref{fig:difficulty}(b) shows that the number of involved files increases significantly as task difficulty rises. However, the number of modified files by all models does not increase correspondingly. This suggests that existing models, including Claude Sonnet 4.5, still exhibit substantial limitations in multi-file coordination.

\subsection{Error Analysis}
\begin{figure}[htbp]
  \centering
\includegraphics[width=\linewidth]
  {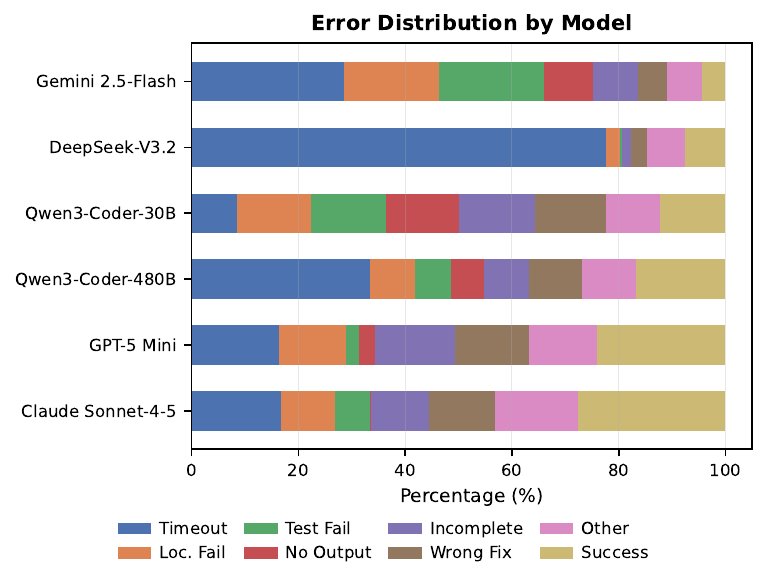}
  \caption{Error distribution of six LLM models on the SmellBench.}
  \label{fig:error_dist}
\end{figure}
This section presents an analysis of model failure cases, which serve as a critical window for understanding model boundaries, facilitating targeted model selection while providing a more comprehensive perspective for the SmellBench evaluation framework.
Specifically, we employed OpenHands as the agent, leveraging DeepSeek V3.2 as a LLM-based classification tool to perform fine-grained categorization of failure cases. This yielded six distinct error categories plus an “Other” category covering sparse error types.
According to Figure \ref{fig:error_dist}, we have the following key findings:
(1) AgentTimeout is a prominent issue. Severe timeout phenomena were observed across all models, reflecting the inherent complexity of the tasks—models require extended reasoning and multi-round revisions. Notably, DeepSeek exhibited a timeout rate as high as 77\%. Further analysis revealed that the model engaged in excessively lengthy reasoning actions, rendering it unable to complete tasks within the limited time.
(2) Bottleneck in localization capabilities. Location Failure accounts for a substantial proportion of errors across all models. This indicates that current models still face challenges in precisely localizing the positions of code smells. 
(3) Insufficient repair and coordination capabilities. Wrong Fix and Incomplete errors both accounted for notable proportions. This suggests that while models have acquired method-level localization capabilities, there still remain substantial room for improvement in formulating concrete repair strategies that how to fix refactor cases and performing cross-file coordinated editing.

\section{Conclusion}
In this paper, we present \textbf{SmellBench}, an extensible benchmark for repository-level code refactoring through proactive code smell injection. SmellBench contains 294 refactoring cases across 7 smell types and introduces a multi-dimensional evaluation framework beyond functional correctness. Experiments on 2 code agents and 6 representative LLMs show that current models still struggle with complex cross-file refactoring tasks, highlighting substantial gaps in repository-level reasoning and software maintainability capabilities.

\section*{Limitations}
We acknowledge several limitations of SmellBench. (1) Since developers may not strictly comply with coding best practices, some code smells, such as God Classes, may naturally exist in the original repositories, which could introduce noise and reduce the reliability of the evaluation results. (2) The current smell injection process exhibits certain similarities across cases, it may encourage models to adopt similar solving patterns, which in turn reduces the diversity of SmellBench. (3) The use of LLM-as-Judge for evaluation introduces potential bias, as the assessment results may be influenced by the capability and inherent preferences of the judging model. This could affect the objectivity and stability of the evaluation outcomes.

\section*{Ethical Impact}

In this paper, we introduce SmellBench, a benchmark designed to evaluate the code refactoring and maintainability capabilities of Code Agents. We recognize and explicitly address the following ethical implications associated with our work:

\paragraph{Data Privacy, Copyright, and Licensing.} 
All source code and software projects included in SmellBench are constructed from [publicly available open-source repositories / synthetically generated pipelines / curated public datasets]. We have strictly adhered to the original licensing agreements (e.g., MIT, Apache 2.0) of all foundational codebeds. We conducted thorough data filtering to ensure that no Personally Identifiable Information (PII), proprietary vendor code, or sensitive data is included in our benchmark.

\paragraph{Code Safety and Dual-Use Risks.} 
While evaluating and enhancing code refactoring capabilities promotes software maintainability and mitigates technical debt, there is a minor risk of "dual-use." Specifically, automated refactoring agents could inadvertently introduce logic flaws, hidden security vulnerabilities (e.g., CWEs), or anti-patterns if deployed in production environments without human supervision. To mitigate this risk, we emphasize that SmellBench is strictly a diagnostic evaluation framework. We strongly advocate that code generated or refactored by LLM-based agents should undergo mandatory human-in-the-loop code review before deployment in mission-critical software systems.
   
\bibliography{custom}

\appendix
\section{Datasets}
Table~\ref{tab:stat_smellbench} summarizes the statistics of the SmellBench benchmark. The benchmark contains 294 cases covering seven widely used open-source Python projects: \texttt{click}, \texttt{jinja}, \texttt{numpy}, \texttt{pandas}, \texttt{scikit-learn}, \texttt{seaborn}, and \texttt{xarray}. On average, each case involves 232.5 lines of code changes, 3.8 files, and 4.3 methods, highlighting the complex multi-file and tightly coupled refactoring characteristics of SmellBench. Moreover, each case contains an average of 23.0 test cases to ensure functional correctness of our SmellBench.

\begin{table}[t]
\centering
\small
\caption{Statistics of SmellBench.}
\label{tab:stat_smellbench}
\begin{tabular}{l|cccc|c}
\toprule
\textbf{Project} & \textbf{Lines} & \textbf{Files} & \textbf{Methods} & \textbf{Tests} & \textbf{Count} \\
\midrule
click & 209 & 3.4 & 3.8 & 24.0 & 42 \\
jinja & 199 & 3.2 & 4.2 & 23.1 & 42 \\
numpy & 241 & 6.9 & 4.1 & 11.4 & 42 \\
pandas & 281 & 3.4 & 4.9 & 37.6 & 42 \\
scikit & 221 & 2.9 & 4.3 & 21.0 & 42 \\
seaborn & 193 & 3.1 & 4.1 & 20.5 & 42 \\
xarray & 284 & 3.6 & 4.5 & 23.6 & 42 \\
\midrule
\textbf{Overall} & \textbf{232.5} & \textbf{3.8} & \textbf{4.3} & \textbf{23.0} & \textbf{294} \\
\bottomrule
\end{tabular}
\end{table}
\label{sec:data_stat}
\section{Token Consumption}
\label{sec:token}
In this section, we discuss the token consumption of our framework. The total token usage consists of two components: generation pipeline consumption and evaluation consumption. As the token cost of the LLM-as-Judge stage is negligible compared to the other components, we omit it from the statistics.

\begin{figure}[htbp]
  \centering
\includegraphics[width=\linewidth]
  {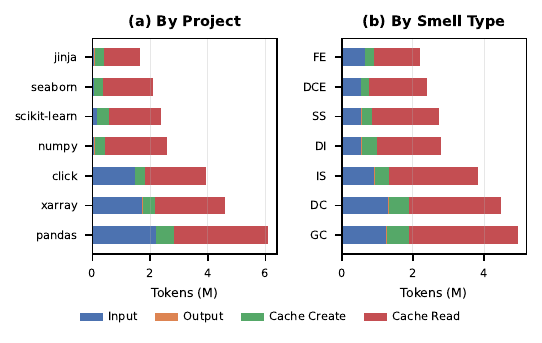}
  \caption{Token consumption in the generation pipeline: (a) grouped by repository; (b) grouped by smell type. }
  \label{fig:token_usage_generation}
\end{figure}
\subsection{Token Consumption in the Generation}
The generation pipeline introduces token consumption in three stages: Candidate Discovery, Smell Injection, and Fixing Errors. All cases in our experiments were generated using Claude Code with Claude Opus 4.6. Detailed statistics for different repositories and smell types are shown below.

Figure \ref{fig:token_usage_generation} shows that generating one case consumes around 2M tokens on average. Most of these tokens are cached tokens, which helps reduce the actual cost significantly. The click, seaborn, and pandas repositories exhibit the highest token consumption. This is mainly because candidate methods in these repositories contain complex dependencies and strong interconnections, requiring multiple synchronized modifications during code transformation.

Similarly, the data clumps smell type often requires modifications across multiple locations, making the identification of related methods highly expensive in terms of tokens. Furthermore, god classes are usually located in large classes or files, where understanding class functionality and dependencies requires substantial contextual reasoning, leading to higher token consumption.

\subsection{Token Consumption in Evaluation}
\begin{figure}[htbp]
  \centering
\includegraphics[width=\linewidth]
  {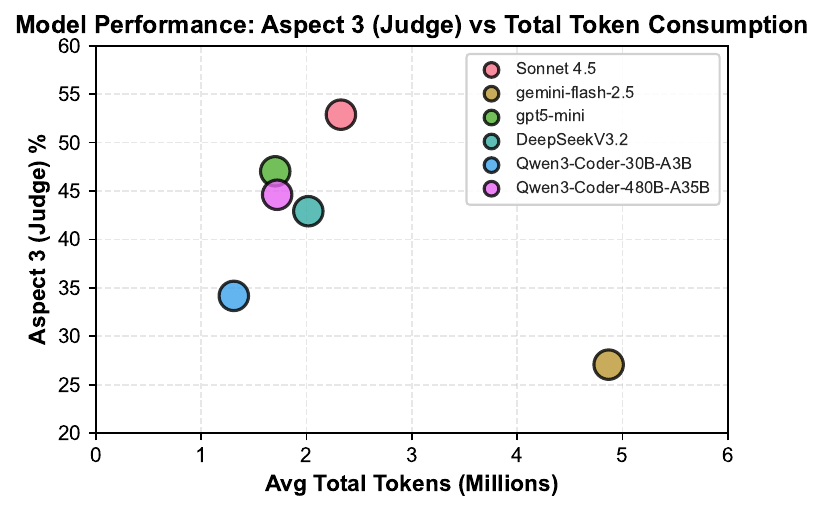}
  \caption{Comparison of model accuracy and efficiency. Higher scores with lower token consumption indicate better efficiency. }
  \label{fig:token_usage_evaluation}
\end{figure}

In this subsection, we further analyze the relationship between model performance and token consumption in order to investigate which models achieve a better trade-off. Based on OpenHands as the evaluation framework, we evaluated the performance and token consumption of six different LLMs. The results are shown in Figure \ref{fig:token_usage_evaluation}.

The experimental results indicate that Sonnet 4.5 achieves the best overall performance among all evaluated models, while maintaining token consumption within an acceptable range, demonstrating strong overall efficiency. In contrast, the open-source models DeepSeek V3.2 and Qwen3-Coder-480B-A35B exhibit performance and token consumption comparable to gpt-5-mini, suggesting that recent high-performance open-source models are increasingly capable of matching the effectiveness and efficiency of proprietary models.

By comparison, Gemini-Flash-2.5 shows relatively weaker overall capability due to its earlier release. It not only consumes more tokens during evaluation, but also significantly underperforms compared with the other models.
\section{Smell Types}
\label{sec:smell_types}
This appendix section describes the code smell types used to construct refactoring cases in our benchmark. Each smell represents a recurring structural deficiency that motivates non-trivial refactoring beyond local edits, often involving interactions across functions, classes, or files.

\subsection{Feature Envy}

\paragraph{Definition.}
Methods that rely heavily on data or behavior from other classes, indicating misplaced responsibilities and poor encapsulation. Such methods are often better located in the classes whose data they primarily manipulate.

\paragraph{Required Capabilities.}
Refactoring Feature Envy requires semantic understanding of responsibility ownership, accurate localization of misplaced methods, and coordinated updates to inter-class dependencies when responsibilities are redistributed across class boundaries.

\subsection{God Classes}

\paragraph{Definition.}
Classes that accumulate excessive responsibilities, resulting in low cohesion, high complexity, and strong coupling with other parts of the system. These classes are typically decomposed into smaller and more cohesive components.

\paragraph{Required Capabilities.}
Addressing God Classes requires architectural reasoning to identify latent responsibility groups, synthesize appropriate class decompositions, and maintain consistency across affected modules after restructuring.

\subsection{Data Clumps}

\paragraph{Definition.}
Groups of variables that frequently appear together across method signatures, constructors, or class interfaces, suggesting missing domain abstractions.

\paragraph{Required Capabilities.}
Refactoring Data Clumps requires recognizing recurring structural patterns, inferring higher-level abstractions, and introducing suitable parameter objects or domain concepts while preserving existing behavior.

\subsection{Shotgun Surgery}

\paragraph{Definition.}
A design problem in which a single conceptual change requires modifications in many scattered locations throughout the codebase, indicating poor localization of responsibilities.

\paragraph{Required Capabilities.}
Eliminating Shotgun Surgery requires repository-scale dependency analysis, accurate identification of all affected locations, and coordinated modifications across multiple files to consolidate change responsibilities.

\subsection{Dead Code Elimination}

\paragraph{Definition.}
The presence of unreachable, redundant, or unused code elements that no longer contribute to program functionality and can be safely removed.

\paragraph{Required Capabilities.}
Refactoring dead code primarily requires reachability analysis, dependency tracking, and behavior-preserving code removal to ensure that no hidden usages or execution paths are disrupted.

\subsection{Interface Segregation}

\paragraph{Definition.}
Overly broad interfaces that force clients to depend on methods they do not use, violating the Interface Segregation Principle and increasing unnecessary coupling.

\paragraph{Required Capabilities.}
Refactoring this smell requires object-oriented design reasoning to identify coherent interface boundaries, decompose large interfaces into focused abstractions, and consistently propagate interface changes to implementations and clients.

\subsection{Deep Inlining}

\paragraph{Definition.}
A refactoring scenario in which deeply nested call chains are collapsed into a single caller, reducing indirection but substantially increasing local implementation complexity while preserving semantics.

\paragraph{Required Capabilities.}
Successfully performing deep inlining requires accurate call-chain analysis, dependency tracking within nested execution paths, and behavior-preserving local code transformations that maintain correctness despite increased complexity.
\section{Difficulty Levels}
\label{sec:difficulty}
Each refactoring case is further stratified into three difficulty levels (\emph{easy}, \emph{medium}, and \emph{hard}) to control structural complexity and refactoring effort. Difficulty is determined by factors such as the number of files involved, and how strongly the smell is obscured by legitimate-looking design patterns. 

As an illustrative example, the difficulty specification for \textbf{Feature Envy} is defined as follows:
\begin{itemize}
    \item \textbf{Easy}: Move one method's core logic to access another class's data directly, spanning exactly two files, with explicit foreign attribute access revealing the envy.
    
    \item \textbf{Medium}: Distribute a method's logic across three or more files via indirect delegation chains (e.g., A calls B which accesses C's internal data), hide the dependency behind wrapper or adapter patterns, and interleave it with legitimate helper methods to introduce diff noise.
    
    \item \textbf{Hard}: Introduce envy through mixin, decorator, or dynamic dispatch patterns so that the cross-class dependency appears intentional; span four or more files with indirect coupling (e.g., callbacks or shared state), include at least one red herring, and require semantic understanding to identify the smell.
\end{itemize}

These difficulty specifications are directly incorporated into the prompt used for case generation. Detailed specifications for all smell types and difficulty levels are provided in \texttt{smell\_types.json} in the \href{https://github.com/MINE-USTC/SmellBench}{https://github.com/MINE-USTC/SmellBench} .
\section{Instruction Settings}
\label{sec:instruction}
We define two instruction settings to evaluate different aspects of model capability: \emph{guided} and \emph{targeted}. Instructions are generated at the instance level, meaning that each refactoring case is associated with a distinct natural-language prompt.

\begin{itemize}
    \item \textbf{Guided}: The instruction specifies only the smell type and the affected file, requiring the model to first localize the smell before performing refactoring.
    
    \item \textbf{Targeted}: The instruction specifies the smell type, file, and the exact class or method involved, focusing the task on executing the refactoring itself. 
\end{itemize}

For example, for case \texttt{b49a19838bb}, the \emph{guided} instruction is:

\begin{quote}
Feature envy has been detected in \texttt{src/click/core.py}. Can you eliminate this smell by moving the logic to a more appropriate location?
\end{quote}

The corresponding \emph{targeted} instruction is:

\begin{quote}
The \texttt{parse\_args} method in the \texttt{Command} class within \texttt{src/click/core.py} exhibits feature envy. Please address this design issue by relocating the behavior to where the data naturally belongs.
\end{quote}

To further investigate how task complexity influences the model's performance under different prompt granularities, we break down the experimental results by difficulty levels (easy, medium, and hard). Table~\ref{tab:difficulty_results} presents the Test Passing Rate (TPR), Localization Accuracy (LA), and average score of LLM-as-Judge (LLM), alongside token consumption.

\begin{table}[htbp]
\centering
\caption{Performance and Token Consumption Across Multi-Level Difficulties. Tokens are measured in Millions}
\label{tab:difficulty_results}
\small 
\setlength{\tabcolsep}{6pt} 
\begin{tabular}{llcccc}
\toprule
\textbf{Setting} & \textbf{Diff} & \textbf{TPR} & \textbf{LA} & \textbf{LLM} & \textbf{Tokens} \\
\midrule
\multirow{3}{*}{Guided} 
 & Easy   & 0.9184 & 0.6735 & 0.4695 & 1.5 \\
 & Med & 0.8776 & 0.6531 & 0.4994 & 2.1 \\
 & Hard   & 0.9796 & 0.6735 & 0.3652 & 1.7 \\
\midrule
\multirow{3}{*}{Targeted} 
 & Easy   & 0.9592 & 0.9184 & 0.5773 & 1.3 \\
 & Med & 0.8776 & 0.9184 & 0.5670 & 1.4 \\
 & Hard   & 0.9592 & 0.7959 & 0.3431 & 1.9 \\
\bottomrule
\end{tabular}
\end{table}

Based on the multi-dimensional breakdown in Table~\ref{tab:difficulty_results}, we obtain several insightful observations:
\begin{itemize}
    \item \textbf{Decoupling of Passing Rate and Quality:} While the TPR remains high ($>0.87$) across all tasks, high functional correctness does not guarantee good refactoring. In the \emph{Guided} setting, despite strong TPR, the LA locates at 0.65--0.67, leading to poor code quality scores from the LLM judge.
    
    \item \textbf{Severe Self-Localization Deficiency:} LLMs exhibit extremely weak autonomous defect localization. In the \emph{Guided} setting, LA stagnates at a low 0.65--0.67 regardless of task difficulty. Providing explicit \emph{Targeted} instruction eliminates this bottleneck, boosting LA up to 0.9184. This contrast highlights a clear performance deficiency of LLMs in identifying code smells autonomously.
    
    \item \textbf{Localization Token Overhead:} Broadly speaking, token consumption scales up with the increase in task difficulty across both settings. However, when examining the horizontal differences, the \emph{Guided} setting does not consistently consume more tokens than \emph{Targeted}, even using fewer tokens in hard tasks (1.7M vs. 1.9M). By analyzing the trajectories, we find that the code agent does not waste extensive steps on localization. Instead, it rapidly settles on a plausible location and proceeds directly to refactoring. This behavior explains both the restricted token costs and the sub-optimal LA under the \emph{Guided} configuration.
\end{itemize}
\section{Rubrics}
\label{sec:rubrics}
In this subsection, we present the scoring criteria for the four rubrics. Specifically, we adopt a 10-point scoring system, where a corresponding description is provided for every two-point interval. For better presentation and readability, Table \ref{tab:overall} converts these scores into a normalized 1-point scale. The detailed criteria are as follows:

\begin{itemize}
    \item \textbf{Code Quality}. Improves code clarity, style, and overall quality to make the code easier to understand and maintain.

 - \textbf{0-1} score: Significantly worse readability than before.

 - \textbf{2-3} score: Poor quality; hard to follow.  

 - \textbf{4-5} score: Below average; multiple readability concerns.

 - \textbf{6-7} score: Acceptable quality; some naming or structural issues.

 - \textbf{8-9} score: Good quality; minor naming or style improvements possible.

 - \textbf{10} score: Clean, idiomatic, well-named, easy to maintain.
 \item \textbf{Structural Soundness}. Maintains a  well-organized architecture with clear module boundaries and reliable design principles

 - \textbf{0-1} score: Introduces new code smells or anti-patterns.

 - \textbf{2-3} score: Significant structural problems.  

 - \textbf{4-5} score: Noticeable structural issues; responsibilities not well separated.

 - \textbf{6-7} score: Reasonable but some unnecessary complexity.

 - \textbf{8-9} score: Sound structure with minor imperfections.

 - \textbf{10} score: Proper decomposition; single responsibility; appropriate abstraction.
 \item \textbf{Cross-File Coordination}. Ensures consistency and correct interaction across multiple files, modules, and dependencies.

 - \textbf{0-1} score: Cross-file coordination largely missing or incorrect.

 - \textbf{2-3} score: Minimal cross-file awareness; related code in other files ignored.

 - \textbf{4-5} score: Some cross-file changes made but several inconsistencies or leftovers.

 - \textbf{6-7} score: Core changes correct; some related artifacts (unused helpers, stale imports) remain in other files.

 - \textbf{8-9} score: Nearly all cross-file impacts handled; one minor leftover.

 - \textbf{10} score: - 10: All smell-related code properly addressed; no orphaned imports, dead helpers, or dangling references left behind.

 \item \textbf{Smell Elimination}. Removes code smells such as duplication, unnecessary complexity, and fragile patterns to improve maintainability.

 - \textbf{0-1} score: Smell not addressed or new smell introduced.

 - \textbf{2-3} score: Minimal effort; smell barely addressed.  

 - \textbf{4-5} score: Only the most obvious smell location fixed; secondary artifacts untouched.

 - \textbf{6-7} score: Core smell addressed but some related artifacts (unused helpers, stale registrations) left behind.

 - \textbf{8-9} score: Smell substantially eliminated; only minor traces remain.

 - \textbf{10} score: Smell fully eliminated; no residual smell code, unused imports, or orphaned helpers remain.
\end{itemize}

\section{Examples}
\begin{figure*}
    \centering
    \includegraphics[width=1\linewidth]{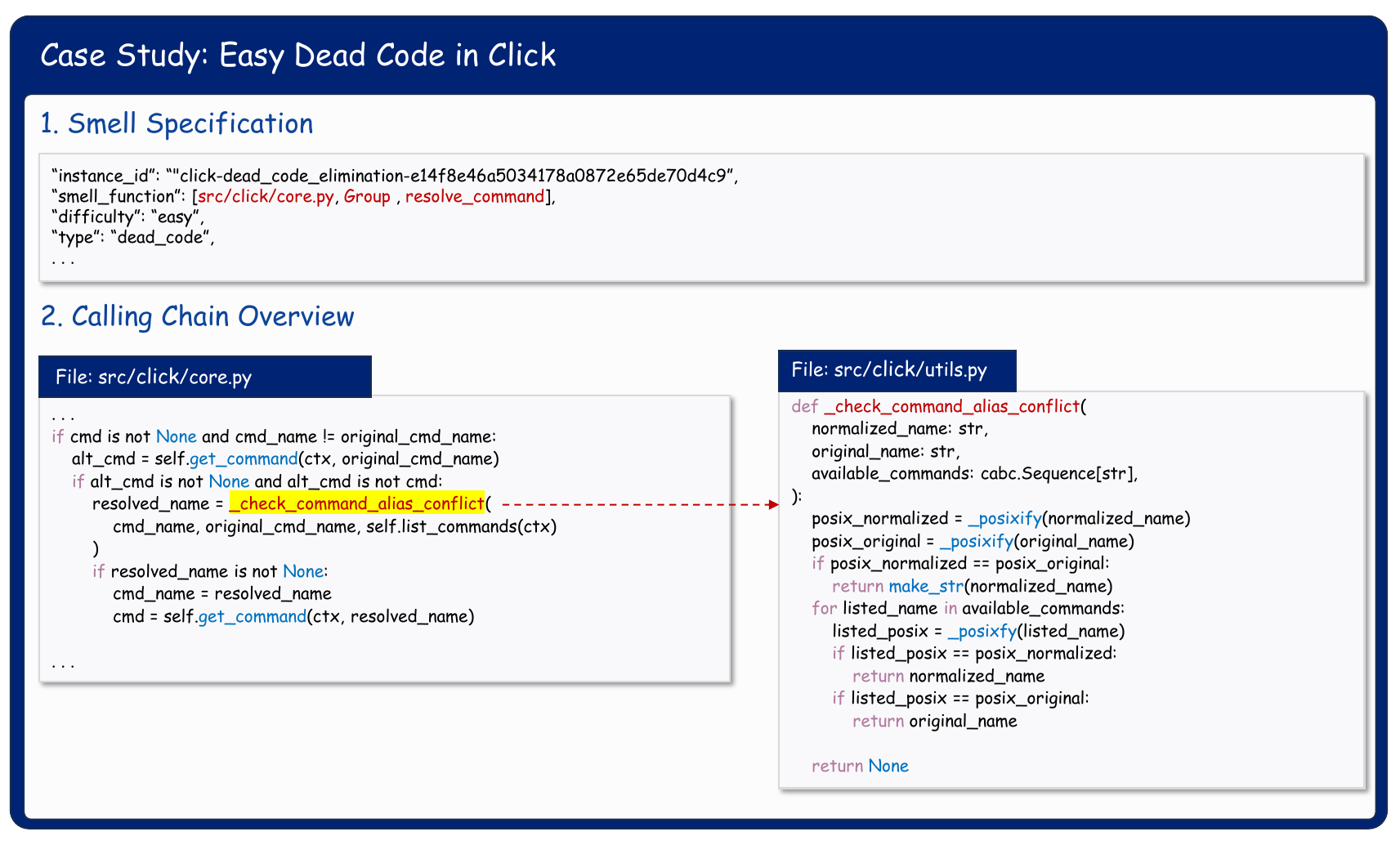}
    \caption{Easy Dead Code Elimination in Click.}
    \label{fig:case_easy}
\end{figure*}

\begin{figure*}
    \centering
    \includegraphics[width=1\linewidth]{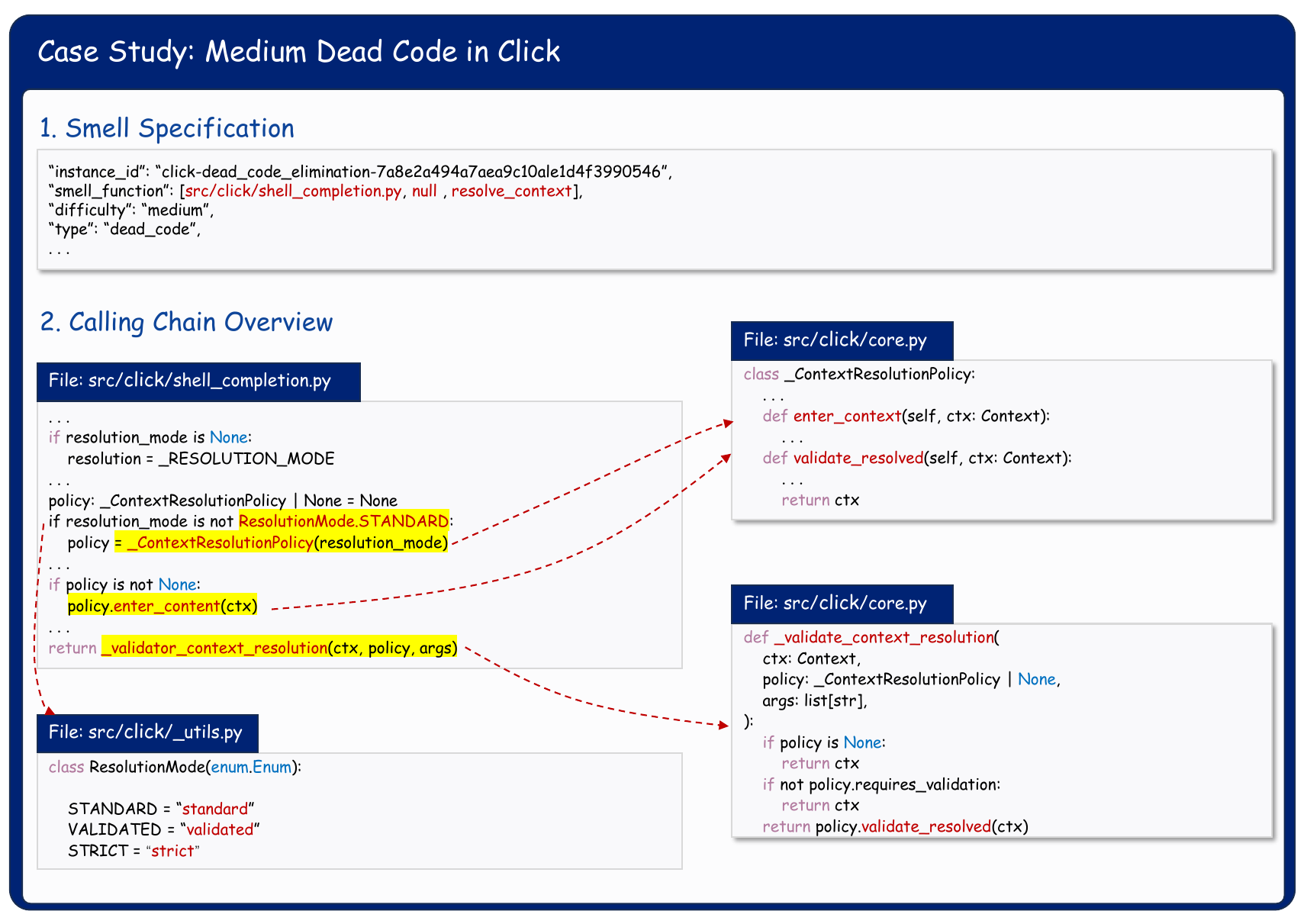}
    \caption{Medium Dead Code Elimination in Click.}
    \label{fig:case_medium}
\end{figure*}

\begin{figure*}
    \centering
    \includegraphics[width=1\linewidth]{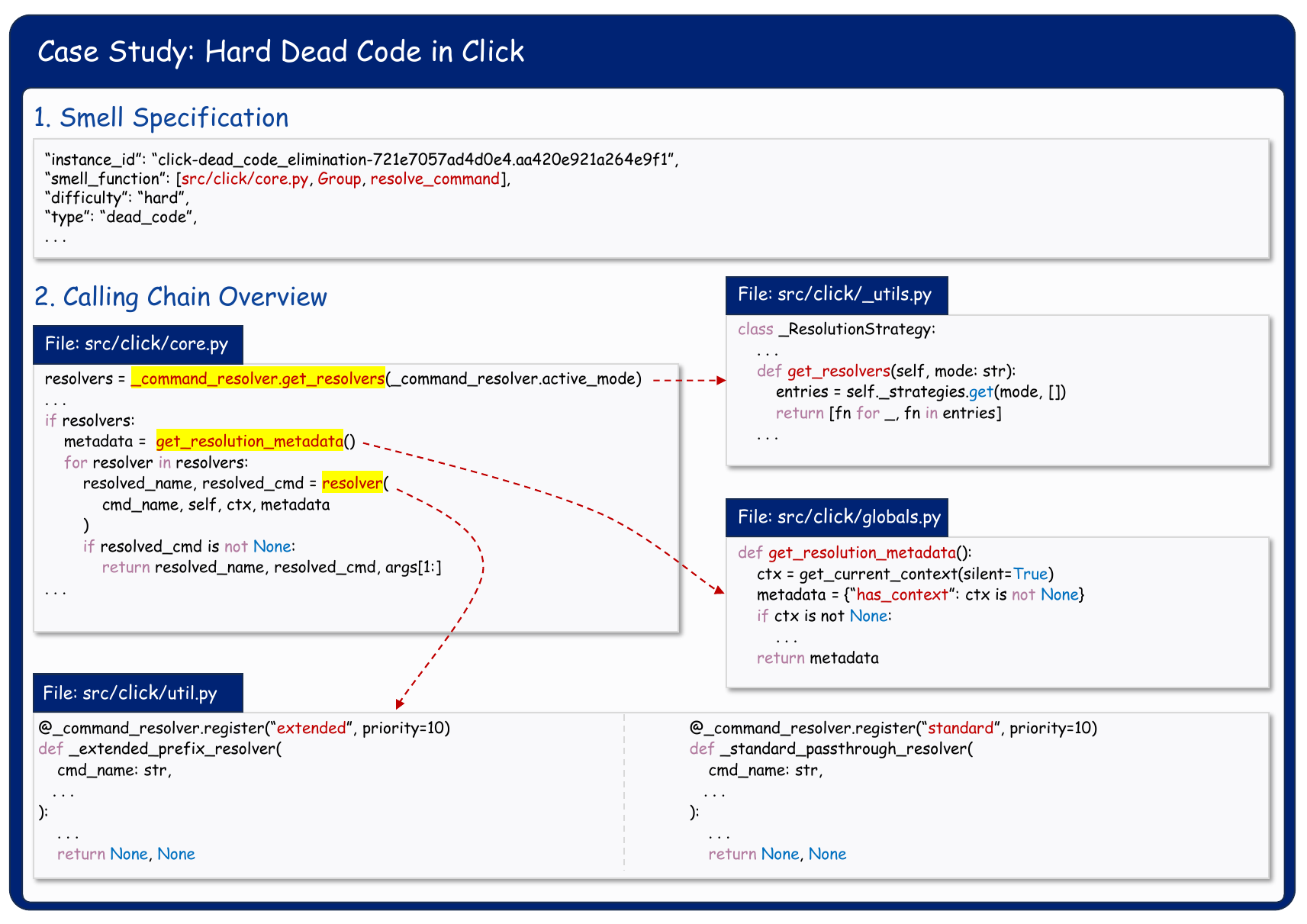}
    \caption{Hard Dead Code Elimination in Click.}
    \label{fig:case_hard}
\end{figure*}

This section presents concrete examples to illustrate how the same smell type differs across difficulty levels. We use the dead code elimination task from the Click project as an example. As shown in Figure \ref{fig:case_easy}, the easy-level task only involves modifications to two files and introduces one new function, \texttt{\_check\_command\_alias\_conflict}, where the dead code appears as an unreachable conditional branch \texttt{resolved\_name is not None}. Figure \ref{fig:case_medium} demonstrates that the medium-level task involves three files, with significantly higher complexity than the easy level. Figure \ref{fig:case_hard} shows that the hard-level task involves four files and introduces the decorator \texttt{\_command\_resolver}, further increasing the concealment of the dead code.

\section{Prompt Design}
\label{sec:prompt}
Below are the prompt templates in our pipelines.

\begin{figure*}[t]
\label{prompt:candidate}
\begin{tcolorbox}[
    colback=white,           
    colframe=gray!80!black,  
    coltitle=white,          
    colbacktitle=gray!80!black, 
    title=Prompt for Candidate Discovery,     
    fonttitle=\bfseries\Large, 
    rounded corners,         
    boxrule=0.5mm           
]

You are a code analysis expert. Your task is to find methods/classes where it would be **convenient to inject** a specific code smell — NOT to find places that already exhibit this smell.
\\
The goal is to identify code locations where the structure, complexity, and cross-module relationships make it natural and easy to introduce the smell while keeping the code compilable and tests passing.
\\
\\
\#\# Source path: `[SRC\_PATH]` 
\\
\\
\#\# Eligible files: [ELIGIBLE\_FILES]
\\
\\
\#\# Smell type to inject: [SMELL\_SPECIFICATION]
\\
\\
\#\# Strategy 

Do NOT read every file. Instead:

1. Based on file names and module structure, pick the 3-5 most promising files

2. Use `grep` to quickly locate class definitions, large methods, and cross-module interactions

3. Only read specific sections of files (use line ranges) to verify candidates

4. Prioritize files with core business logic (e.g., core.py, models.py, engine.py) over peripherals
\\
\\
\#\# Requirements

Find exactly 5 candidates. Each candidate should be a method/class where injecting `[SMELL\_TYPE]` would be **easy and natural** — meaning the surrounding code structure supports the injection without breaking functionality.
\\
\\
Each candidate needs:

- `file`: relative path from repo root

- `class\_name`: class name (null if standalone function)

- `method\_name`: method/function name (for god\_classes/interface\_segregation, can be null)

- `line\_number`: the actual starting line number (verify by reading)

- `reason`: 1-2 sentences explaining why this location is a good **injection point** (what structural properties make it easy to introduce the smell here)
\\
\\
**IMPORTANT**: At most 2 candidates may come from the same file. Spread candidates across different files to ensure diversity.
\\
\\
**DIVERSITY REQUIREMENT**: The 5 candidates must be substantially different from each other:

- They should involve different classes/functions with different responsibilities

- They should target different code patterns or architectural concerns

- Avoid picking multiple methods from the same class or methods that do similar things
\\
\\
\#\# Output
After finding all candidates, output a single JSON block:

\begin{lstlisting}[language=json, numbers=none]
{
  "[SMELL_TYPE]": [
    {"file": "...", "class_name": "...", "method_name": "...", "line_number": 123, "reason": "..."}
  ]
}
\end{lstlisting}
The key must be exactly: [SMELL\_TYPE]

It must have exactly 5 entries.
\end{tcolorbox}

\caption{Candidate discovery prompt illustrating the discovery process.}
\label{fig:candidate_discovery_prompt}
\end{figure*}

\begin{figure*}[t]
\begin{tcolorbox}[
    colback=white,           
    colframe=gray!80!black,  
    coltitle=white,          
    colbacktitle=gray!80!black, 
    title=Prompt for Smell Injection,     
    fonttitle=\bfseries\Large, 
    rounded corners,         
    boxrule=0.5mm           
]
You are a senior software engineer specializing in code transformation and refactoring benchmarks.

Your task is NOT to improve the code. Instead, you must inject a specific type of code smell into the given codebase.
\\
\\
\#\#\# Objective \\
Inject the following code smell:
[SMELL\_TYPE] \\
into the target location: 
[PROJECT\_PATH] 
\\
\\
\#\#\# Target Candidate

You MUST inject the smell starting from the following candidate:
[CANDIDATE PATH]
\\
\\
\#\#\# Difficulty Level: [DIFFICULTY]
[DIFFICULTY\_DESC]
\\
\\
\#\#\# Constraints 

[Constraints Details]
\\
\\
\#\#\# Output Requirements

After generating the modified codebase, you MUST append a JSON object at the very end of your response.
Format:
\begin{lstlisting}[language=json, numbers=none]
{
  "smell_type": "Type name",
  "hint_targeted": "...",
  "hint_guided": "...",
  "smell_function": ["absolute/path/to/file.java", "ClassName", "methodName"],
  "test_functions": [
    ["absolute/path/to/file.java", "ClassName", "methodName"],
    ["absolute/path/to/other.java", "ClassName", "methodName"]
  ]
}
\end{lstlisting}

\#\#\# smell\_function rules

[Smell Function DESC]
\\
\\
\#\#\# test\_functions rules

[TEST FUNCTION DESC]
\end{tcolorbox}

\caption{Prompt for Smell Inject, which is the main component of our construction pipeline.}
\label{fig:smell_injection_prompt}

\end{figure*}

\begin{figure*}[t]
\begin{tcolorbox}[
    colback=white,           
    colframe=gray!80!black,  
    coltitle=white,          
    colbacktitle=gray!80!black, 
    title=Prompt for Fixing Errors,     
    fonttitle=\bfseries\Large, 
    rounded corners,         
    boxrule=0.5mm           
]

You previously injected a "[SMELL\_TYPE]" code smell into this codebase.

The changes you made caused the following unit tests to FAIL.
\\
\\
**CRITICAL**: Your task is to fix the TESTS, NOT to remove the smell!
\\
\\
**Your task**: Fix the code so that all tests pass while 

**KEEPING the injected code smell intact**.

- The smell must still be present and non-trivial after your fix

- You are fixing **test failures caused by your injection**, NOT refactoring the smell away

- Think of it as: "make my smell injection more robust so it doesn't break tests"
\\
\\
**DO NOT**:

- Remove or refactor the smell you injected

- "Clean up" the code smell

- Make the code better quality — keep the smell as is
\\
\\
**DO**:

- Fix any syntax errors, import errors, or runtime errors

- Adjust test-related code if needed

- Make minimal changes to pass tests while preserving the smell pattern
\\
\\
Do NOT run any tests yourself — testing is handled externally.
\\
\\
\#\# Diff of your previous changes

[SMELL\_CONTENT]
\\
\\
\#\# Failing test scripts

[TEST\_SCRIPTS]
\\
\\
\#\# Test error output

[TEST\_ERROR\_OUTPUT]
\\
\\
\#\# Requirements

1. **Fix the failing tests while preserving the code smell injection** — the smell pattern must remain

2. The code must compile/run correctly

3. **DO NOT remove or refactor the smell** — you are fixing test failures, not improving code quality

4. DO NOT create new files

5. DO NOT run any test commands (pytest, unittest, etc.)

6. Make **minimal changes** — only fix what's broken, keep the smell as you originally injected it

After making your fixes, output the same JSON format as before:

\begin{lstlisting}[language=json, numbers=none]
{
  "smell_type": "Type name",
  "hint_targeted": "...",
  "hint_guided": "...",
  "smell_function": ["absolute/path/to/file", "ClassName", "methodName"],
  "test_functions": [
    ["absolute/path/to/file", "ClassName", "methodName"],
    ["absolute/path/to/other", "ClassName", "methodName"]
  ]
}

\end{lstlisting}
\end{tcolorbox}
\caption{Prompt for fixing errors from Quality Validation module .}
\label{fig:fixing_error_prompt}
\end{figure*}

\begin{figure*}[t]
\begin{tcolorbox}[
    colback=white,           
    colframe=gray!80!black,  
    coltitle=white,          
    colbacktitle=gray!80!black, 
    title=Prompt for Smell Analysis,     
    fonttitle=\bfseries\Large, 
    rounded corners,         
    boxrule=0.5mm           
]

You are an expert software engineer analyzing a code change (diff) that introduces a "{smell\_type}" code smell into a codebase.
\\
\\
\#\# Smell Type

- **Name**: {smell\_type}

- **Description**: {smell\_description}

\#\# The Smell Diff

This diff was applied to a clean codebase to introduce the smell:

[smell\_diff]
\\
\\
\#\# Your Task

Analyze this diff by going through each change and assessing its significance. Focus on **what each change means and how important it is**, NOT on how to fix it.

For each distinct change in the diff (a new function, a moved block, an added import, etc.), explain:

- **What it does**

- **How significant it is** (critical / moderate / minor) to the smell

- **What it degrades** in the codebase (e.g., coupling, cohesion, readability, API surface, etc.)
\\
\\
After covering individual changes, provide:

1. **Overall smell pattern**: Summarize how these changes work together to create the "{smell\_type}" smell. What design principle is violated?

2. **Severity ranking**: Rank the changes from most to least important. Which changes are the **root cause** of the smell, and which are just supporting noise?

3. **What was degraded overall**: What concrete qualities of the codebase were harmed? Be specific about the impact on maintainability, coupling, cohesion, etc.

4. **Key evaluation signals**: When judging whether a candidate fix truly addresses this smell, what should matter most? What would distinguish a thorough fix from a superficial one?
\\
\\
\#\# Output Format

Return your result using XML tags. Do NOT wrap the output in a code block.

<analysis>

Your full analysis text as described above. Write freely — no escaping needed.

</analysis>
\end{tcolorbox}
\caption{Prompt for smell analysis to produce detailed information to facilitate comprehensive evaluation.}
\label{fig:smell_analysis_prompt}
\end{figure*}
\begin{figure*}[t]
\begin{tcolorbox}[
    colback=white,           
    colframe=gray!80!black,  
    coltitle=white,          
    colbacktitle=gray!80!black, 
    title=Example Output for Smell Analysis,     
    fonttitle=\bfseries\Large, 
    rounded corners,         
    boxrule=0.5mm           
]
\#\#\# Change 1: Storing `\_last\_state` in `\_OptionParser` (parser.py)

**What it does**: Adds a single line `self.\_last\_state = state` to persist the parsing state object after parsing completes. This exposes internal parsing state that was previously encapsulated within the parse method.
\\
\\
**Significance**: CRITICAL - This is the enabling infrastructure change that makes the feature envy possible. Without this, the Command class couldn't access the parser's internal state.
\\
\\
**What it degrades**: 
\\
- **Encapsulation**: Breaks the parser's encapsulation by exposing internal state (`\_ParsingState`) that should remain private
- **API surface**: Implicitly expands the parser's interface by making `\_last\_state` a queryable attribute
\\
\\
- **Coupling**: Creates a temporal coupling - the `\_last\_state` is only valid after parsing, but there's no contract enforcing this
\\
...
\end{tcolorbox}
\caption{Example output of smell analysis used in LLM-as-Judge.}
\label{fig:smell_analysis_example}
\end{figure*}

\begin{figure*}[t]
\begin{tcolorbox}[
    colback=white,           
    colframe=gray!80!black,  
    coltitle=white,          
    colbacktitle=gray!80!black, 
    title=Prompt for Task Instruction,     
    fonttitle=\bfseries\Large, 
    rounded corners,         
    boxrule=0.5mm           
]
You are an expert Python refactoring assistant.

\#\# Task

[guided or targeted instruction]
\\
\\
\#\#\# Instructions

1. Read and understand the relevant code in the project.

2. Identify the code smell and understand why it is problematic.

3. Refactor the code to eliminate the smell while preserving all existing behavior.

4. Ensure all tests continue to pass after your changes.
\\
\\
\#\#\# Constraints
\\
- Do not change the original program behavior.

- Preserve the original control flow, inputs, and outputs.
\\
\\
\#\#\# IMPORTANT

You are allowed to perform any operations, such as checking or modifying related files, without prior approval from me.
\\
\\
\#\#\# Response

\end{tcolorbox}
\caption{Task instruction prompt provided to the refactoring agent. }
\label{fig:task_instruction}

\end{figure*}

\begin{figure*}[t]
\begin{tcolorbox}[
    colback=white,           
    colframe=gray!80!black,  
    coltitle=white,          
    colbacktitle=gray!80!black, 
    title=Prompt for LLM-as-Judge,     
    fonttitle=\bfseries\Large, 
    rounded corners,         
    boxrule=0.5mm           
]
You are an expert code reviewer evaluating a refactored version of code that originally contained a "[smell\_type]" code smell.
\\
\\
\#\# Context
\\
- **Smell Type**: [smell\_type]
\\
- **Smell Description**: [description]

[smell\_analysis]
\\
\\
**IMPORTANT**: The refactoring below is provided in **unified diff format** (git diff output). Lines starting with `-` are removed, lines starting with `+` are added, and context lines are unchanged. Evaluate the *intent and quality of the changes*, not the completeness of the code shown — diffs only show changed regions, not the full files.
\\
\\
\#\#\# Refactoring (diff fixing the smell)

[refactored\_code]
\\
\\
\#\# Evaluation Rubric

[general\_dimensions]
\\
\\
Return your evaluation as JSON:

\begin{lstlisting}[language=json, numbers=none]
{
  "smell_elimination": {
    "score": 0,
    "justification": "brief explanation"
  },
  "cross_file_coordination": {
    "score": 0,
    "justification": "brief explanation"
  },
  "structural_soundness": {
    "score": 0,
    "justification": "brief explanation"
  },
  "code_quality": {
    "score": 0,
    "justification": "brief explanation"
  },
  "summary": "2-3 sentence overall assessment"
}
\end{lstlisting}
\end{tcolorbox}
\caption{Judging Prompt for LLM to evaluate the refactoring result of agents.} 
\end{figure*}

\end{document}